\documentclass[twocolumn,prd,superscriptaddress,showpacs,floatfix,%
preprintnumbers,nofootinbib]{revtex4}

\usepackage{epsfig}

\begin{document}

\title{Self-induced spectral splits in supernova neutrino fluxes}

\author{Georg G.~Raffelt}
\affiliation{Max-Planck-Institut f\"ur Physik
(Werner-Heisenberg-Institut), F\"ohringer Ring 6, 80805 M\"unchen,
Germany}

\author{Alexei Yu.~Smirnov}
 \affiliation{Max-Planck-Institut f\"ur Physik
 (Werner-Heisenberg-Institut), F\"ohringer Ring 6, 80805 M\"unchen,
Germany} \affiliation{Abdus Salam International Centre for
Theoretical Physics, Strada Costiera 11, 34014 Trieste, Italy}
\affiliation{Institute for Nuclear Research, Russian Academy of
Sciences, 117 312 Moskva, Russia}

\date{12 May 2007, revised 17 July 2007, corrected 7 Dec 2007}

\preprint{MPP-2007-53}

\begin{abstract}
In the dense-neutrino region above the neutrino sphere of a supernova
($r\alt 400$~km), neutrino-neutrino refraction causes collective
flavor transformations. They can lead to ``spectral splits'' where an
energy $E_{\rm split}$ splits the transformed spectrum sharply into
parts of almost pure but different flavors. Unless there is an
ordinary MSW resonance in the dense-neutrino region, $E_{\rm split}$
is determined by flavor-lepton number conservation alone. Spectral
splits are created by an adiabatic transition between regions of large
and small neutrino density. We solve the equations of motion in the
adiabatic limit explicitly and provide analytic expressions for a
generic example.
\end{abstract}


\maketitle


\section{Introduction}

At large densities, neutrino-neutrino refraction
causes nonlinear flavor oscillation phenomena with sometimes
perplexing results~\cite{Pantaleone:1992eq, Samuel:1993uw,
  Samuel:1996ri, Qian:1995ua, Fuller:2005ae, Pastor:2001iu,
  Pastor:2002we, Duan:2005cp, Duan:2006an, Duan:2006jv,
  Hannestad:2006nj, Duan:2007mv, Mirizzi2007, Raffelt:2007yz,
  EstebanPretel:2007ec}. In the region between the neutrino sphere and
a radius of about 400~km in core-collapse supernovae (SNe), the
neutrino flavor content evolves dramatically~\cite{Pastor:2002we,
  Duan:2005cp, Duan:2006an, Duan:2006jv, Hannestad:2006nj,
  Duan:2007mv, Mirizzi2007}. The global features of this self-induced
transformation are equivalent to the motion of a gyroscopic pendulum
in flavor space~\cite{Hannestad:2006nj, Duan:2007mv}. However, this
picture does not explain the ``spectral splits'' that have been
numerically observed in the transformed fluxes~\cite{Duan:2006an,
  Duan:2006jv, Mirizzi2007}. In a typical case, the primary $\nu_e$
flux below a split energy $E_{\rm split}$ emerges from the
dense-neutrino region in its original flavor, whereas above $E_{\rm
  split}$, it is completely transformed to $\nu_x$ (some mixture of
$\nu_\mu$ and $\nu_\tau$), the step at $E_{\rm split}$ being very
sharp. (To be specific we explore the $\nu_e$--$\nu_x$ system with the
atmospheric $\Delta m^2$ and the small 13-mixing angle.)

It has been suggested that an adiabatic transition from high to
low neutrino density is the primary cause for the
split~\cite{Duan:2006an, Duan:2007mv}. Dense neutrinos perform
synchronized oscillations: all modes oscillate with a common
frequency $\omega_{\rm synch}$, even though their individual
frequencies vary as $\omega =  |\Delta m^2/2E|$. Flavor oscillations
can be visualized as the precession of polarization vectors ${\bf
P}_\omega$ in a ``flavor ${\bf B}$ field.'' The ${\bf P}_\omega$
``stick together'' by the $\nu$--$\nu$--interaction, thus forming a
collective object that precesses around~${\bf B}$. The collectivity
is lost when the neutrino density decreases. However, if the
decrease is slow, all ${\bf P}_\omega$ align themselves with or
against ${\bf B}$ in the process of decoupling from each other.
Eventually they all precess with their individual $\omega$ around
${\bf B}$, but without visible consequences because of their
(anti-)alignment with~${\bf B}$.

We extend this interpretation of the split phenomenon in several
ways. We (i)~show that flavor-lepton number conservation determines
$E_{\rm split}$, (ii)~solve the equations of motion explicitly in the
adiabatic limit, and (iii)~provide an analytic result for a generic
case.

\section{Equations of motion}

We represent the flavor content of an
isotropic $\nu$--$\bar\nu$ gas by flavor polarization vectors ${\bf
P}_\omega$ and $\bar{\bf P}_\omega$, where overbarred quantities
correspond to $\bar\nu$. We define their global counterparts as
${\bf P}=\int_0^\infty d\omega\,{\bf P}_\omega$ and $\bar{\bf
P}=\int_0^\infty d\omega\,\bar{\bf P}_\omega$ and introduce ${\bf D}
\equiv {\bf P}-\bar{\bf P}$, representing the net lepton number. The
equations of motion (EOMs) are~\cite{Hannestad:2006nj, Sigl:1992fn}
\begin{equation}\label{eq:eom1}
 \partial_t{\bf P}_\omega=\left(\omega{\bf B}+\lambda{\bf L}
 +\mu {\bf D}\right)\times{\bf P}_\omega
\end{equation}
and the same for $\bar{\bf P}_\omega$ with $\omega\to-\omega$. Here
$\lambda\equiv\sqrt{2}G_F n_e$ represents the usual matter potential
and $\mu \equiv \sqrt{2}G_F n_\nu$ the $\nu$--$\nu$ interaction
strength, where $n_e$ and $n_\nu$ are the electron and neutrino
densities. We work in the mass basis where ${\bf B}=(0,0,-1)$
corresponds to the normal and ${\bf B}=(0,0,+1)$ to the inverted mass
hierarchies. The interaction direction ${\bf L}$ is a unit vector such
that ${\bf B}\cdot{\bf L}=\cos2\theta$ with $\theta$ being the vacuum
mixing angle. Unless there is an MSW resonance in the dense-neutrino
region, one can eliminate $\lambda{\bf L}$ from Eq.~(\ref{eq:eom1}) by
going into a rotating frame, at the expense of a small effective
mixing angle~\cite{Duan:2005cp, Hannestad:2006nj}. The only difference
for antineutrinos is that in vacuum they oscillate ``the other way
round.'' Therefore, instead of using $\bar{\bf P}_\omega$ we may
extend ${\bf P}_\omega$ to negative frequencies such that $\bar{\bf
  P}_{\omega} = {\bf P}_{-\omega}$ ($\omega > 0$) and use only ${\bf
  P}_\omega$ with $-\infty<\omega<+\infty$. In these terms, ${\bf D}=
\int_{-\infty}^{+\infty}d\omega\,s_\omega\,{\bf P}_\omega$, where
$s_\omega \equiv {\rm sign}(\omega)=\omega/|\omega|$.

After elimination of  $\lambda{\bf L}$, the EOM for ${\bf D}$ can be
obtained by integrating Eq.~(\ref{eq:eom1}) with  $s_\omega$:
\begin{equation}
\label{eq:eomD}
 \partial_t{\bf D} = {\bf B} \times{\bf M}
 \hbox{\quad where \quad}
 {\bf M} \equiv
 \int_{-\infty}^{+\infty} d\omega\, s_\omega \omega {\bf P}_\omega\,.
\end{equation}
It shows that  $\partial_t({\bf D}\cdot{\bf B})=0$ so that $D_z
= {\bf B}\cdot{\bf D}$ is conserved~\cite{Hannestad:2006nj}. The
in-medium mixing angle above a SN core is small and therefore the
mass and interaction basis almost coincide. Collective effects then
only induce pair transformations of the form
$\nu_e\bar\nu_e\to\nu_x\bar\nu_x$, whereas the excess $\nu_e$ flux
from deleptonization is conserved.

\section{Adiabatic solution}

We rewrite the EOMs in terms of an
``effective Hamiltonian'' for the individual modes as
\begin{equation}
\label{eq:eom2}
 \partial_t{\bf P}_\omega={\bf H}_\omega\times{\bf P}_\omega
 \hbox{\quad where\quad}
 {\bf H}_\omega = \omega{\bf B} +\mu {\bf D}.
\end{equation}
In the adiabatic limit each ${\bf H}_\omega$ moves slowly compared
to the precession of ${\bf P}_\omega$ so that the latter follows the
former. We assume that initially all ${\bf P}_\omega$ represent the
same flavor and thus are aligned. If initially $\mu$ is large, every
${\bf P}_\omega$ is practically aligned with ${\bf H}_\omega$.
Therefore, in the adiabatic limit it stays aligned with ${\bf
H}_\omega$ for the entire evolution:
\begin{equation}\label{eq:hialigned}
 {\bf P}_\omega (\mu) = \hat{\bf H}_\omega(\mu)\,P_\omega\,,
\end{equation}
which solves the EOMs. Here $P_\omega \equiv |{\bf P}_\omega|$ and
$\hat{\bf H}_\omega \equiv {\bf H}_\omega/|{\bf H}_\omega|$ is a unit
vector. Here and henceforth we assume an excess flux of neutrinos over
antineutrinos, implying that initially ${\bf P}_\omega$ and ${\bf D}$
are collinear and $D_z>0$.

According to Eq.~(\ref{eq:eom2}) all ${\bf H}_\omega$ lie in the
plane spanned by ${\bf B}$ and ${\bf D}$  which we call the
``co-rotating plane.'' In the adiabatic limit  all ${\bf P}_\omega$,
and consequently ${\bf M}$, also stay in that plane.
Therefore we can decompose
\begin{equation}\label{decompose}
 {\bf M} = b\, {\bf B} + \omega_{\rm c} {\bf D}
\end{equation}
and rewrite the EOM of Eq.~(\ref{eq:eomD}) as
\begin{equation}
\label{eq:eomD2}
\partial_t{\bf D} = \omega_{\rm c}\,{\bf B} \times{\bf D}.
\end{equation}
Therefore ${\bf D}$ and the co-rotating plane precess around ${\bf
B}$ with the common or ``co-rotation frequency'' $\omega_{\rm c}$.

We conclude that the system evolves simultaneously in two ways: a
fast precession around ${\bf B}$ determined by $\omega_{\rm c} =
\omega_{\rm c}(\mu)$ and  a drift in the co-rotating plane caused by
the explicit $\mu(t)$ variation. To isolate the latter from the
former, we go (following Ref.~\cite{Duan:2005cp}) into the
co-rotating frame where the individual Hamiltonians become
\begin{equation}\label{eq:Hi}
 {\bf H}_\omega=(\omega-\omega_{\rm c})\,{\bf B}
 +\mu {\bf D}\,.
\end{equation}
We use the same notation because the relevant components $H_{\omega
z}$, $H_{\omega \perp}$, $D_{z}$, and $D_{\perp}$ remain invariant.

Initially ($\mu\to\infty$) the oscillations are synchronized,
$\omega_{\rm c}^{\infty}=\omega_{\rm synch}$, and all ${\bf P}_\omega$
form a collective~${\bf P}$. As $\mu$ decreases, the ${\bf P}_\omega$
zenith angles spread out while remaining in a single co-rotating
plane. In the end ($\mu\to 0$) the co-rotation frequency is
$\omega_{\rm c}^0$ and Eqs.~(\ref{eq:hialigned}) and~(\ref{eq:Hi})
imply that all final ${\bf H}_\omega$ and therefore all ${\bf
  P}_\omega$ with $\omega > \omega_{\rm c}^0$ are aligned with ${\bf
  B}$, the others anti-aligned: a spectral split is inevitable with
$\omega_{\rm split}\equiv\omega_{\rm c}^0$ being the split
frequency. The lengths $P_\omega=|{\bf P}_\omega|$ are conserved and
eventually all ${\bf P}_\omega$ point in the $\pm{\bf B}$
directions. Therefore the conservation of flavor-lepton number gives
us $\omega_{\rm split}$, for $D_z>0$, by virtue of
\begin{equation}\label{spl}
 D_z=\int_{-\infty}^0 P_\omega\,d\omega
 -\int_{0}^{\omega_{\rm split}}P_\omega\,d\omega
 +\int_{\omega_{\rm split}}^{+\infty} P_\omega\,d\omega\,.
\end{equation}
In general, $\omega_{\rm split}=\omega_{\rm c}^0 \neq \omega_{\rm
c}^\infty=\omega_{\rm synch}$.

For individual modes the EOMs given by ${\bf H}_\omega$ are
completely solved if we find $\omega_{\rm c}(\mu)$ and
$D_\perp(\mu)$, the component transverse to ${\bf B}$, since $D_z$
is conserved and given by the initial condition. From
Eq.~(\ref{eq:hialigned}) we infer $P_{\omega\perp}/P_\omega
 =H_{\omega\perp}/H_\omega$, from Eq.~(\ref{eq:Hi})
$H_{\omega\perp}=\mu D_\perp$ and $H_{\omega z}=\omega-\omega_{\rm
c}+\mu D_z$ so that
\begin{eqnarray}
 P_{\omega,z}&=&\frac{(\omega-\omega_{\rm c}+\mu D_z)\,P_\omega}
 {\sqrt{(\omega-\omega_{\rm c}+\mu D_z)^2+(\mu D_\perp)^2}}\,,
 \label{eq:pz}\\*
 P_{\omega\perp}&=&\frac{\mu D_\perp\,P_\omega}
 {\sqrt{(\omega-\omega_{\rm c}+\mu D_z)^2+(\mu D_\perp)^2}}\,.
 \label{eq:pperp}
\end{eqnarray}
Integration of the second equation over $s_\omega d\omega$ gives us
\begin{equation}\label{eq:master1}
 1=\int_{-\infty}^{+\infty}d\omega\,s_\omega\,\frac{P_\omega}
 {\sqrt{[(\omega-\omega_{\rm c})/\mu+D_z]^2+D_\perp^2}}\,.
\end{equation}
Projecting Eq.~(\ref{decompose}) on the $x$--$y$--plane we find
$\omega_{\rm c} = M_{\perp} / D_\perp(\mu)$ or explicitly
\begin{equation}\label{eq:Dfreq}
 \omega_{\rm c}=
 \frac{\int_{-\infty}^{+\infty} d\omega\,s_\omega\,\omega\,P_{\omega\perp}}
 {\int_{-\infty}^{+\infty} d\omega\,s_\omega P_{\omega\perp}}
 =\frac{\int_{-\infty}^{+\infty} d\omega\,s_\omega\,\omega\,P_{\omega\perp}}
 {D_\perp}\,.
\end{equation}
For large $\mu$ when the oscillations are synchronized, this agrees
with the usual expression for $\omega_{\rm synch}$
\cite{Pastor:2001iu}, but it changes when the ${\bf P}_\omega$
spread out in the zenith direction. Inserting Eq.~(\ref{eq:pperp})
into Eq.~(\ref{eq:Dfreq}) we find
\begin{equation}\label{eq:master2}
 \omega_{\rm c}=\int_{-\infty}^{+\infty} d\omega\,s_\omega\,
 \frac{\omega\,P_\omega}
 {\sqrt{[(\omega-\omega_{\rm c})/\mu+D_z]^2+D_\perp^2}}\,.
\end{equation}
Given $D_z$ and a spectrum $P_\omega$, we can determine $\omega_{\rm
  c}$ and $D_\perp$ from Eqs. (\ref{eq:master1}) and
(\ref{eq:master2}) for any $\mu$.  These equations solve the EOMs
explicitly in the adiabatic limit.

We have assumed that all ${\bf P}_\omega$ are initially aligned.
One can relax this restriction and allow  some ${\bf P}_\omega$ to
have opposite orientation. If different species are emitted from a
SN core with equal luminosities but different average energies, the
spectra will cross over so that some range of modes is prepared,
say, as $\nu_e$ and another as $\nu_x$.

\begin{figure*}
\includegraphics[width=0.9\textwidth]{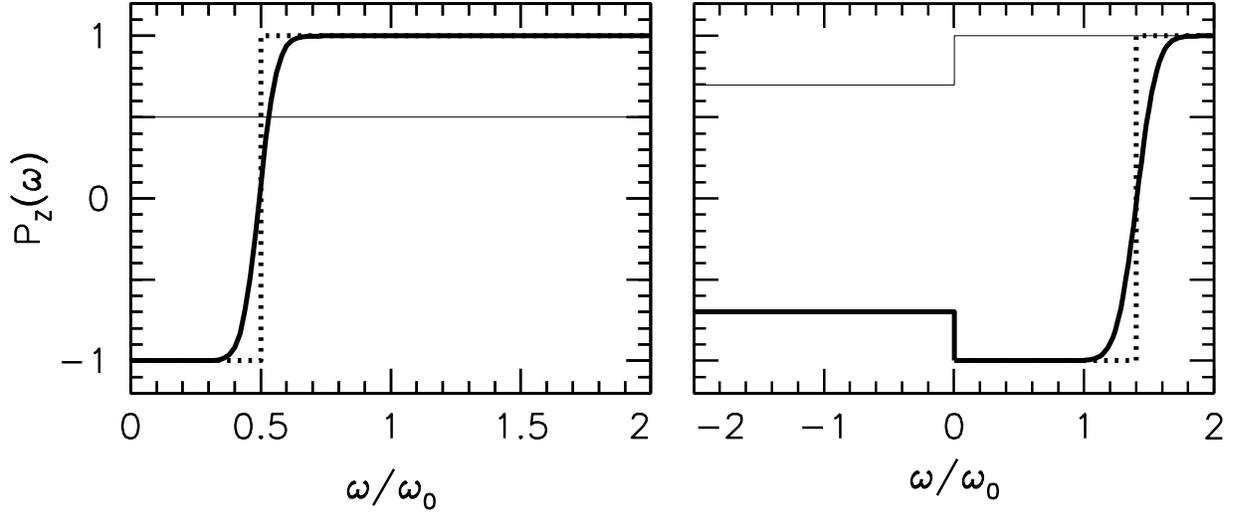}
\caption{Spectra of polarization vectors ($z$-component). Thin:
initial. Thick: final. Dotted: fully adiabatic. Solid: numerical
solution as described in the text. {\it Left:} Box-like initial
$\nu$ spectrum, large misalignment between ${\bf B}$ and ${\bf P}$,
and no~$\bar\nu$. {\it Right:} Box-like $\nu$ and $\bar\nu$ spectra,
30\% fewer $\bar\nu$, small initial misalignment
($\sin2\theta=0.05$), and inverted
hierarchy.\label{fig:boxspectrum}}
\end{figure*}
\begin{figure*}
 \includegraphics[width=0.9\textwidth]{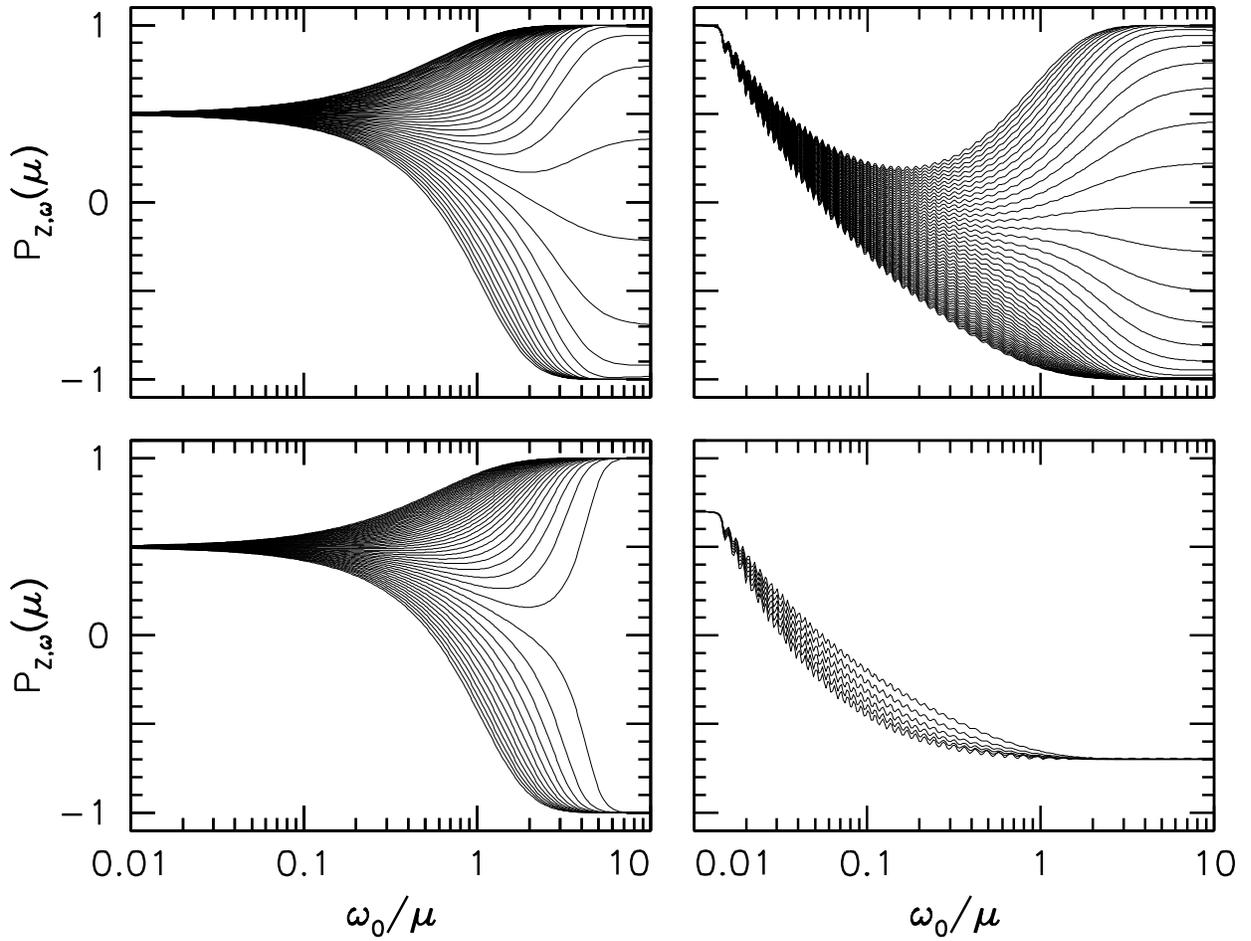}
\caption{$P_{z,\omega}(\mu)$ for 51~modes. {\it Left:} Box-like
$\nu$-only spectrum. Numerical solution of EOMs ({\it top}).
Analytic adiabatic solution ({\it bottom}). {\it Right:} Box-like
$\nu$ and $\bar\nu$ spectra. Numerical solution for $\nu$ ({\it
top}) and $\bar\nu$ ({\it bottom}), here only 6
modes.\label{fig:spectrumevolution}}
\end{figure*}

\section{Neutrinos only}

We illustrate the power of our new results
with a generic neutrino-only example (${\bf D}={\bf P}$). The
spectrum is taken box like with $P_\omega=(2\omega_0)^{-1}$ for
$0\leq\omega\leq2\omega_0$ and~0 otherwise.  With $P_z$ being
conserved we find from Eq.~(\ref{spl})
\begin{equation}\label{eq:fsplit}
 \omega_{\rm c}=\omega_0\times\cases{1&for $\mu\to\infty$,\cr
 (1-P_z)&for $\mu\to0$.\cr}
\end{equation}
The case $P_z=0$ is special because $\omega_{\rm c}=\omega_0$
remains fixed. For $P_z=1$ we have $\omega_{\rm c}^0=0$ and no
flavor evolution. We use $P_z=0.5$ to show the initial and final
$P_z(\omega)$ in Fig.~\ref{fig:boxspectrum} (left). The dotted line
denotes the adiabatic final state where $\omega_{\rm split} =
0.5\,\omega_0$. The solid line is from a numerical solution of the
EOMs with $\mu(t)=\mu_0\,\exp(-t/\tau)$ and
$\tau^{-1}=0.03\,\omega_0$, typical for a~SN. We have checked
numerically that the split indeed becomes sharper with increasing
$\tau$ and thus increasing adiabaticity.

In Fig.~\ref{fig:spectrumevolution} we show $P_{\omega,z}(\mu)$ for
51 individual modes. They start with the common value $P_{\omega,z}
= 0.5\,(2\omega_0)^{-1}$. Later they spread and eventually split,
some of them approaching $+1$ and the others $-1$. Some modes first
move down and then turn around as $\omega_{\rm c}$ changes. A few
modes do not reach $\pm1$ because of imperfect adiabaticity.

For the box spectrum the integrals Eqs.~(\ref{eq:master1})
and~(\ref{eq:master2}) are easily performed and one can extract
\begin{eqnarray}\label{eq:exact1}
 \omega_{\rm c}&=&\omega_0+
 \omega_0 P_z\left(\frac{1}{\kappa} -
 \frac{e^{\kappa}+e^{-\kappa}}{e^{\kappa}-e^{-\kappa}}\right)\,,
 \nonumber\\*
 P_\perp&=&\sqrt{1-P_z^2}\;
 \frac{2\kappa}{e^{\kappa}-e^{-\kappa}}\,,
\end{eqnarray}
where $\kappa \equiv \omega_0/\mu$. For $\mu\to\infty$ and $\mu\to0$
the limits of $\omega_{\rm c}$ agree with Eq.~(\ref{eq:fsplit}) from
lepton-number conservation. For $\mu\to\infty$ we obtain
$P_\perp=\sqrt{1-P_z^2}$, representing the initial condition $P=1$,
and for $\mu\to0$ we find $P_\perp=0$.

With Eq.~(\ref{eq:pz}) these results provide analytic solutions for
the adiabatic $P_{\omega,z}(\mu)$. We show examples in
Fig.~\ref{fig:spectrumevolution} (bottom left) for comparison with
the numerical solution of the EOMs. The agreement is striking and
confirms the picture of adiabatic evolution in the co-rotating
plane. The agreement is poor for modes close to the split ($\omega
\approx \omega_{\rm c}^0$) at low neutrino densities ($\mu <
\omega_0$) where the evolution becomes nonadiabatic.

\section{Adiabaticity condition}

The speed for the  ${\bf H}_\omega$
evolution in the co-rotating plane is $d\theta_\omega/dt$, where
$\cos \theta_\omega \equiv H_{\omega\perp}/H_\omega$, while ${\bf
P}_\omega$ precesses with speed~$H_\omega$. The evolution is
adiabatic if the adiabaticity parameter $\gamma_\omega \equiv
|d\theta_\omega/dt|\,H_\omega^{-1}\ll1$. With Eqs.~(\ref{eq:Hi})
and~(\ref{eq:pperp}) we find
\begin{equation}\label{gamma}
 \gamma_\omega=
 \frac{\displaystyle
 \left(\frac{\omega-\omega_{\rm c}}{\mu}
 +D_z\right)\frac{dD_\perp}{d\mu} +
 \frac{D_\perp}{\mu}\frac{d\omega_{\rm c}}{d\mu} +
 D_\perp\frac{\omega-\omega_{\rm c}}{\mu^2}}
 {\tau_\mu\displaystyle
 \left[\left(\frac{\omega-\omega_{\rm c}}{\mu}
 +D_z\right)^2 + D_\perp^2 \right]^{3/2}}\,,
\end{equation}
where $\tau_\mu\equiv|d\ln\mu/dt|^{-1}$.

For our neutrino-only ($D_{\perp} \rightarrow P_{\perp}$) box
spectrum  Eqs.~(\ref{eq:exact1}) give
$dP_\perp/d\mu=-P_\perp(\omega_{\rm c} -\omega_0)/\mu^2$ and
$d\omega_{\rm c}/d\mu=P_z[1-4\kappa^2/(e^\kappa-e^{-\kappa})^2]$.
For $\mu \gg \omega_0$ we obtain $dP_\perp/d\mu\sim
\omega_0^2/\mu^3$ and $d\omega_{\rm c}/d\mu \sim \omega_0^2/\mu^2$
so that the last term in the numerator of Eq.~(\ref{gamma})
dominates: $\gamma_{\omega} \sim P_\perp(\omega-\omega_{\rm
c})/(h_\mu \mu^2)$. With $\mu$ decreasing, $\gamma_{\omega}$
increases and at $\mu \sim \omega_0$ when $\gamma_{\omega} \sim 1$,
adiabaticity violation begins. For $\mu < \omega_0$ the denominator
of Eq.~(\ref{gamma}) gives the dependence $\gamma_{\omega} \propto
(\omega-\omega_{\rm c})^{-3}$, and therefore the closer $\omega$
to~$\omega_{\rm c}$ the stronger the adiabaticity violation.

\section{Including antineutrinos.}

As a second generic case we now add
antineutrinos. One important difference is that even a very small
initial misalignment between ${\bf D}$ and ${\bf B}$ is enough to
cause a strong effect. Consider a single energy mode for $\nu$ with
$P=1$ and one for $\bar\nu$ with $\bar P=\alpha<1$ that are
initially aligned in the flavor direction, now taken very close to
the mass direction, and assume an inverted hierarchy. From the
dynamics of the flavor pendulum~\cite{Hannestad:2006nj, Duan:2007mv}
we know that in the end $\bar {\bf P}$ is antialigned with ${\bf
B}$, whereas ${\bf P}$ retains a large transverse component because
$P_z-\bar P_z$ is conserved: The system prepares itself for a
spectral split.

Assuming box spectra for both $\nu$ and $\bar\nu$, we show the
initial and final $P_{z,\omega}$ in Fig.~\ref{fig:boxspectrum}
(right), for the inverted hierarchy, $\sin2\theta=0.05$, and
$\alpha=0.7$. From Eq.~(\ref{eq:Dfreq}) one infers $\omega_{\rm
c}^\infty=\omega_{\rm synch}=
 \omega_0\,(1+\alpha)/(1-\alpha)$.
For $\alpha=\frac{7}{10}$ this is $\omega_{\rm c}^\infty=
\frac{17}{3}\,\omega_0>2\omega_0$. Therefore, all modes have
negative frequencies in the co-rotating frame and tilt away from
${\bf B}$ (see also the numerical $P_{\omega,z}$ in
Fig.~\ref{fig:spectrumevolution}). The final split frequency is
found from flavor lepton number conservation to be $\omega_{\rm split}
=\omega_0\,(1 - D_z + \alpha)\approx\omega_0\, 2\alpha$, using
$D_z\approx1-\alpha$ for $\sin2\theta\ll1$. With
$\alpha=\frac{7}{10}$ we find $\omega_{\rm split}=
\frac{14}{10}\,\omega_0$ in agreement with
Fig.~\ref{fig:boxspectrum}. For $0<\alpha<1$ we have $0<\omega_{\rm
split}<2\omega_0$ so that the final split always occurs among the
neutrinos. According to~Fig.~\ref{fig:spectrumevolution} the split
starts when the vector {\bf D} develops a significant transverse
component, and it proceeds  efficiently in a region  $\mu \sim
\omega_0$.

The ``wiggles'' in the curves in the right panels of
Fig.~\ref{fig:spectrumevolution} stem from the nutation of the flavor
pendulum~\cite{Hannestad:2006nj, Duan:2007mv}.  We have chosen a
relatively fast $\mu(t)$ evolution ($\tau^{-1}=0.1\,\omega_0$),
implying poor adiabaticity, to avoid too many nutation periods on the
plot. For a very slow $\mu(t)$ the nutations disappear and the
co-rotating frame removes the full global evolution of the system.

\section{Discussion}

We have studied the phenomenon of spectral
splits that is caused by neutrino-neutrino refraction in the SN
dense-neutrino region. We have carried previous explanations of this
novel effect~\cite{Duan:2006an, Duan:2007mv} to the point of
explicit solutions in the adiabatic limit.

A spectral split occurs when a neutrino ensemble is prepared such
that the common direction of the flavor polarization vectors
deviates from the mass direction. An adiabatic density decrease
turns all modes below a split energy $E_{\rm split} \equiv \Delta
m^2/2\omega_{\rm split}$ into the mass direction, and the others in
the opposite direction. Remarkably, during this phase all modes
remain in a single rotating plane, even after losing full
synchronization. $E_{\rm split}$ is determined by lepton number
conservation in the mass basis.

The spectral split is a generic feature of the adiabatic evolution
when the density changes from large to small values.  It can appear
even in the absense of neutrino-neutrino interactions.  Indeed, in the
usual MSW case the evolution to zero density transforms $\nu_e$ to
$\nu_2$ and  $\bar{\nu}_e$ to $\bar{\nu}_1$ for all
energies. This corresponds to $\omega_{\rm split} = 0$. The
neutrino-neutrino interactions shift $\omega_{\rm split}$ to non-zero
values.

A spectral split is caused in the SN neutrino (but not antineutrino)
flux by neutrino-neutrino interactions alone, especially during the
accretion phase when ordinary MSW resonances occur far outside the
dense-neutrino region.  Later the matter profile may become so shallow
that the H-resonance moves into this region~\cite{Duan:2006an,
  Duan:2006jv, Duan:2007mv}. The simultaneous action of collective
effects and an ordinary MSW resonance may then cause spectral splits
for both neutrinos and antineutrinos, leading to a rich phenomenology,
perhaps modifying r-process nucleosynthesis~\cite{Duan:2006an,
  Duan:2007mv}.  Of course, the fluxes will be further processed by
ordinary conversion in the SN envelope~\cite{Dighe:1999bi,
  Dighe:2004xy}, thus modifying observable signatures.  Still,
observing spectral splits would provide a smoking gun signature both
for the relevant neutrino properties and, if it occurs among
antineutrinos at late times, for the occurrence of a shallow density
profile above the neutrino sphere.

The neutrino flux emitted by a SN is anisotropic so that
neutrinos on different trajectories experience different
neutrino-neutrino interaction histories~\cite{Duan:2006an,
  Duan:2007mv} that would be expected to cause kinematical flavor
decoherence of different angular modes~\cite{Raffelt:2007yz}. A
numerical exploration reveals, however, that in a typical SN scenario
the deleptonization flux suppresses decoherence and the evolution is
almost identical to that of an isotropic
ensemble~\cite{EstebanPretel:2007ec}. Our treatment of the spectral
evolution is apparently applicable in a realistic SN context.

Collective neutrino oscillation phenomena in a SN may well be
important for the explosion mechanism, r-process nucleosynthesis and
may provide detectable signatures in a high-statistics signal from the
next galactic SN. Building on previous ideas, our formalism gives a
simple, elegant and quantitative explanation of seemingly impenetrable
numerical results.  Our approach provides the basis for developing a
quantitative understanding of realistic consequences of collective
neutrino oscillations for SN physics and observational signatures.


\begin{acknowledgments}
We acknowledge support by the Deutsche Forschungsgemeinschaft (TR~27
``Neutrinos and beyond''), the European Union (ILIAS project,
contract RII3-CT-2004-506222), the Alexander von Humboldt
Foundation, and The Cluster of Excellence ``Origin and Structure of
the Universe'' (Munich and Garching).
\end{acknowledgments}



\begin{thebibliography}{00}

\bibitem{Pantaleone:1992eq}
  J.~Pantaleone,
  ``Neutrino oscillations at high densities,''
  Phys.\ Lett.\ B {\bf 287}, 128 (1992).

\bibitem{Samuel:1993uw}
  S.~Samuel,
  ``Neutrino oscillations in dense neutrino gas\-es,''
  Phys.\ Rev.\ D {\bf 48}, 1462 (1993).

\bibitem{Samuel:1996ri}
  S.~Samuel,
  ``Bimodal coherence in dense selfinteracting neutrino gases,''
  Phys.\ Rev.\ D {\bf 53}, 5382 (1996)
  [hep-ph/ 9604341].

\bibitem{Qian:1995ua}
  Y.~Z.~Qian and G.~M.~Fuller,
  ``Matter enhanced anti-neutrino flavor transformation
  and supernova nucleosynthesis,''
  Phys.\ Rev.\  D {\bf 52}, 656 (1995)
  [astro-ph/9502080].

\bibitem{Fuller:2005ae}
  G.~M.~Fuller and Y.~Z.~Qian,
  ``Simultaneous flavor transformation of neutrinos and antineutrinos
  with dominant potentials from neutrino neutrino forward
  scattering,''
  Phys.\ Rev.\ D {\bf 73}, 023004 (2006)
  [astro-ph/0505240].

\bibitem{Pastor:2001iu}
  S.~Pastor, G.~G.~Raffelt and D.~V.~Semikoz,
  ``Physics of synchronized neutrino oscillations caused by
  self-interactions,''
  Phys.\ Rev.\ D {\bf 65}, 053011 (2002)
  [hep-ph/0109035].

\bibitem{Pastor:2002we}
  S.~Pastor and G.~Raffelt,
  ``Flavor oscillations in the supernova hot bubble region:
  Nonlinear  effects of neutrino background,''
  Phys.\ Rev.\ Lett.\  {\bf 89}, 191101 (2002)
  [astro-ph/0207281].

\bibitem{Duan:2005cp}
  H.~Duan, G.~M.~Fuller and Y.~Z.~Qian,
  ``Collective neutrino flavor transformation in supernovae,''
  Phys.\ Rev.\  D {\bf 74}, 123004 (2006)
  [astro-ph/0511275].

\bibitem{Duan:2006an}
  H.~Duan, G.~M.~Fuller, J.~Carlson and Y.~Z.~Qian,
  ``Simulation of coherent non-linear neutrino flavor transformation
  in the supernova environment. I: Correlated neutrino trajectories,''
  Phys.\ Rev.\ D {\bf 74}, 105014 (2006)
  [astro-ph/0606616].

\bibitem{Duan:2006jv}
  H.~Duan, G.~M.~Fuller, J.~Carlson and Y.~Z.~Qian,
  ``Coherent development of neutrino flavor in the supernova
  environment,''
  Phys.\ Rev.\ Lett.\ {\bf 97}, 241101 (2006)
  [astro-ph/0608050].

\bibitem{Hannestad:2006nj}
  S.~Hannestad, G.~G.~Raffelt, G.~Sigl and Y.~Y.~Y.~Wong,
  ``Self-induced conversion in dense neutrino gases: Pendulum
  in flavour space,''
  Phys.\ Rev.\ D {\bf 74}, 105010 (2006)
  [astro-ph/0608695].

\bibitem{Duan:2007mv}
  H.~Duan, G.~M.~Fuller, J.~Carlson and Y.~Z.~Qian,
  ``Analysis of collective neutrino flavor transformation
  in supernovae,''
  arXiv:astro-ph/0703776.

\bibitem{Mirizzi2007}
  G.~L.~Fogli, E.~Lisi, A.~Marrone and A.~Mirizzi,
  ``Collective neutrino flavor transitions in supernovae and the role of
  trajectory averaging,''
  arXiv:0707.1998 [hep-ph].

\bibitem{Raffelt:2007yz}
  G.~G.~Raffelt and G.~Sigl,
  ``Self-induced decoherence in dense neutrino gases,''
  Phys.\ Rev.\ D {\bf 75}, 083002 (2007)
  [hep-ph/0701182].

\bibitem{EstebanPretel:2007ec}
  A.~Esteban-Pretel, S.~Pastor, R.~Tom\`as, G.~G.~Raffelt and G.~Sigl,
  ``Decoherence in supernova neutrino transformations suppressed by
  deleptonization,''
  arXiv:0706.2498 [astro-ph].

\bibitem{Sigl:1992fn}
  G.~Sigl and G.~Raffelt,
  ``General kinetic description of relativistic mixed neutrinos,''
  Nucl.\ Phys.\ B {\bf 406}, 423 (1993).

\bibitem{Dighe:1999bi}
  A.~S.~Dighe and A.~Y.~Smirnov,
  ``Identifying the neutrino mass spectrum from the neutrino
  burst from a supernova,''
  Phys.\ Rev.\  D {\bf 62}, 033007 (2000)
  [hep-ph/9907423].

\bibitem{Dighe:2004xy}
  A.~Dighe,
  ``Supernova neutrinos: Production, propagation and oscillations,''
  Nucl.\ Phys.\ Proc.\ Suppl.\  {\bf 143}, 449 (2005)
  [hep-ph/0409268].

\end{thebibliography}
\end{document}